\def\etal{{\it et al.\/}}

\def\ie{{\it i.e.\/}}

\documentstyle[aps,prl,floats]{revtex}
\begin{document}
\draft
\tighten
\twocolumn[\hsize\textwidth\columnwidth\hsize\csname@twocolumnfalse%
\endcsname
\title{$GeV$ Photons from Ultra High Energy Cosmic Rays accelerated in
Gamma Ray Bursts}
\author{Mario Vietri}
\address{Dipartimento di Fisica E. Amaldi, Universit\`a di Roma 3
\\ via della Vasca Navale 84, 00147 Roma 
\\ E-mail: vietri@corelli.fis.uniroma3.it}
\date{\today}
\maketitle
\begin{abstract}
$\gamma$-ray bursts are produced by the dissipation of the kinetic energy of a 
highly relativistic fireball, via the formation of a collisionless shock. When 
this happens, Ultra High Energy Cosmic Rays up to $\approx10^{20}\;eV$ are 
produced. I show in this paper that these particles produce, via synchrotron 
emission as they cross the acceleration region, photons up to $300\; GeV$
which carry away a small, $\approx 0.01$, but non--negligible 
fraction of the total burst energy. I show that, when the shock occurs
with the interstellar medium, the optical depth to photon--photon scattering,
which might cause energy degradation of the photons, is small. The burst 
thusly produced would
be detected at Earth simultaneoulsy with the parent $\gamma$--ray burst,
although its duration may differ significantly from that of the lower energy
photons. The expected fluences, $\approx 10^{-5}-10^{-6}\; erg 
\; cm^{-2}$, are well within the range of planned detectors. 
A new explanation for the exceptional burst {\it GB 940217} is discussed. 
\end{abstract}

]

\narrowtext

It can be argued compellingly that a model for extragalactic $\gamma$--ray
bursts (GRBs) must involve highly relativistic motions \cite{review}. This model
(fireball) involves the acceleration of matter to large Lorenz factors 
($\gamma \approx 100-1000$) from a compact object, and the formation of a 
highly relativistic shock. The shock converts the
directed kinetic energy (mostly of the baryons) into internal energy,
equally distributed between electrons and protons. Electrons then
promptly radiate their share of the internal energy, giving rise, through
synchrotron radiation in a suitably generated magnetic field, to the
observed GRB. 

The relativistic environment surrounding the above--mentioned shock
seems suitable for the acceleration of protons to high energies \cite{vietri}. 
In non--relativistic shocks, particle acceleration is a painfully slow
process: particles shuffle diffusively from downstream back to upstream
and viceversa, each time increasing their speed infinitesimally (this
is the modern version of the Fermi mechanism, \cite{fermi}). Instead,
in relativistic shocks, the distribution function of non--thermal
particles in the fluid frame is strongly collimated in the direction 
perpendicular to the shock, and they suffer deflections which differ little 
from forward/backward scattering \cite{quenby}. 
Furthermore, at each cross shocking, their energies are {\it 
multiplied}\/ by the factor $\gamma^2$; for the large Lorenz factors for the 
shocks in the GRBs' scenarios described above, the largest known in the 
Universe, just two or three cycles suffice to propel
protons to energies $\approx 10^{20}\; eV$. In GRBs, a large part of the energy 
loss must go through this channel, but the mechanism seems so fast and powerful 
that it has been suggested that the whole flux of Ultra High Energy Cosmic Rays 
(UHECRs) observed at Earth is generated in GRBs \cite{vietri,wax1}.

The major rival to acceleration of UHECRs in GRBs is acceleration in blazars,
or in the hot spots of radio galaxies \cite{proth}. This mechanism is however 
hampered by the paucity of blazars and radio galaxies inside the 
Greizen--Zatsepin--Kuzmin limit \cite{gzk}, $< 100\; 
Mpc$. Thus it has problems explaining the rough isotropy of the directions of
arrival of the UHECRs, and the lack of any suitable candidate as the site of
acceleration around the direction of arrival of the highest energy cosmic ray 
observed so far \cite{elbert}, which are not a problem for the GRB theory 
\cite{waxreview}.

However, the GRB scenario ought to find independent confirmation. 
One possible test has been proposed already \cite{wax2}: it 
involves the fact that the spectrum of cosmic rays emitted by a {\it single}
source, as observed at the Earth, is wide (a power--law) if the source is a
continuous emitter, but is much narrower if the source is explosive, because of
energy--dependent time--delays in the propagation of cosmic rays in the magnetic
field of intergalactic regions. In this {\it Letter}, I propose a different test
which may also reveal something about the details of the acceleration process. 

\paragraph*{Synchrotron emission by UHECRs.}

In models for cosmological GRBs, two different scenarios for 
the generation of the shock have been envisioned so far: in the first
\cite{laguna}, a shock is generated when the ejecta crash into 
the interstellar medium, much like a SuperNova.
In the second \cite{rees94}, the compact object generates two expanding 
relativistic shells, endowed with slightly different
Lorenz factors; when the second, faster shell overcomes the first one,
collisionless shocks propagate through both shells. It has been argued
\cite{sari} that the first mechanism may be responsible for bursts with
smooth lightcurves, and the second one for spiky lightcurves. For reasons to be
explained later, I concentrate on the first one. 

In this scenario, a total energy release $E_{GRB} = E_{51} 10^{51} \; erg$ is 
contaminated with a baryon mass $M_b$ such that 
\begin{equation}
\label{baryons}
\eta\equiv E_{GRB}/M_b c^2 \approx 10^3\;.
\end{equation}
The baryons, and the Coulomb--dragged electrons, are accelerated to a
Lorenz factor $\gamma \approx \eta$, and then start a coasting phase
which is terminated with the formation of a shock with the interstellar medium, 
of number density $n_1 \approx 1 \; cm^{-3}$, for typical 
ISM environments. The shock forms approximately when a total ISM matter
$\approx M/\eta$ has been collected. This occurs at a radius
\begin{equation}
r_{d} = 10^{18} \; cm\; E_{51}^{1/3} n_1^{-1/3} \eta^{-2/3}\;.
\end{equation}
At this moment, the just--formed shock splits: a forward shock moves
into the ISM matter, with Lorenz factor $\sqrt{2}\eta$, while a reverse
shock propagates backward into the ejecta, causing the shell's deceleration.
It can be shown \cite{laguna} that the ejecta shell thickness at this moment
is given by $r_{sh} = r_d/\eta$ in the shell frame, where the 
post--shock magnetic field is $B = 1\; G \;n_1^{1/2} \eta \xi^{1/2}$,
with $\xi$ parametrizing departures from exact equipartition (corresponding 
to $\xi = 1$); for this GRB model to work properly, it is necessary that
that $\xi \approx 1$, which I shall assume henceforth.

It was argued in \cite{vietri} that the distribution of non--thermal
protons extends from $E_l \approx \eta^2 m_p c^2 \approx 10^{15} \; eV$ to 
\begin{equation}
E_u = 10^{19} \; eV \; \eta^{1/3} E_{51}^{1/3} n_1^{1/6} \xi^{1/2}\;,
\end{equation}
all energies measured at the Earth. In the shell frame, the lower and upper 
Lorenz factors become $\gamma_l \approx \eta$ and $\gamma_u \approx
10^8 (\eta/10^3)^{-2/3} E_{51}^{1/3} n_1^{1/6} \xi^{1/2}$.

The spectrum of non--thermal particles behind non--relativistic shocks
is a power--law with index $p\approx2$. 
Albeit less is known about relativistic shocks, a similar conclusion
holds \cite{shocks}. This is consistent with observations:
Waxman \cite{wax3} has computed the injection spectrum at GRBs, such that,
after inclusion of photopion losses \cite{gzk} and cosmological effects,
these particles fit the cosmic ray spectrum observed at Earth, for observed
energies exceeding $3\times 10^{18}\; eV$. He finds good agreement
for any index such that $1.8 < p < 2.3$, consistent with production of
UHECRs in GRBs. I shall thus take, in the fluid frame, 
$d\!n = N_\circ \gamma^{-p} \; d\!\gamma$, for $\gamma_l < \gamma < \gamma_u$. 

For a power--law energy distribution of non--thermal particles, the spectrum 
of synchrotron emission follows a power--law $\propto \nu^{-s}$ with index
$s = (p-1)/2$, $ \approx 0.5$ in our case. Thus synchrotron emission is
heavily dominated by the high--energy end of the spectrum. Each non--thermal
particle, in the fluid frame, emits at a typical frequency given by 
$\omega_c = 3\epsilon^2 e B/2 m_p^3 c^5$, where I dropped the inessential
dependence on the angle between the field and the particle mean velocity,
and $\epsilon$ is the particle energy in the fluid frame $\approx E/\eta$. 
The high--energy end of the photons' spectrum is cutoff at the typical 
emission frequency $\omega_{uc}$ of the highest energy particles, $E_u$, 
which, scaling to the values given above, and trasforming the emission 
frequency to the Earth's reference frame, is
\begin{equation}
\label{cutoff}
E_{co} = \eta \hbar \omega_{uc} = 1 \; GeV \eta^{2/3} E_{51}^{2/3}
n_1^{-1/6} \xi^{3/2}\;;
\end{equation}
for the values favoured by observations of $\eta = 10^3$ and $E_{51} = 4$ 
\cite{piran}, the cutoff energy becomes $E_{co} \approx 300 \; GeV$.

While the spectral shape and the cutoff energy are easy to derive, the
overall normalization is trickier. In non--relativistic shocks \cite{volk}
the total energy emitted as radiation is of the same order of magnitude
as that channeled into non--thermal particles; while for relativistic
shocks no equivalent computation exists, still the arguments given above make 
it clear that relativistic shocks ought to be, if anything, more efficient than 
non--relativistic ones in accelerating protons. Assuming rough equipartition, 
the total flux of UHECRs at Earth is explained by the GRBs' scenario 
\cite{vietri,wax1}. Thus I will equate the total energy released in cosmic rays
$E_{CR}$ to that released by the burst in radiation, both, of course,
in the shell frame:
\begin{equation}
\label{ecr}
E_{CR} \equiv N_\circ m_p c^2 \int_{\gamma_l}^{\gamma_u} \gamma^{-1} \;
d\!\gamma = \frac{E_{GRB}}{\eta}\;, 
\end{equation}
obtaining $N_\circ = E_{GRB}/\eta m_p c^2 \ln\gamma_u/\gamma_l$. 

The total mass in non--thermal particles $M_{CR}$ thusly determined is 
reassuringly small: we have
\begin{equation}
M_{CR} = N_\circ m_p \int_{\gamma_l}^{\gamma_u} \gamma^{-2} \; d\!\gamma = 
\frac{N_\circ m_p}{\gamma_l} \;,
\end{equation}
which, using Eqs. \ref{baryons}, \ref{ecr} and $\gamma_l \approx \eta$, gives
\begin{equation}
\frac{M_{CR}}{M_b} = \frac{1}{\eta \ln\gamma_u/\gamma_l} \ll 1\;.
\end{equation}

The synchrotron energy loss per particle per unit time is given by
$\dot{\epsilon} = - 2 e^4 B^2 \epsilon^2/ (3 m^4 c^7)$. 
Most of the energy is lost by the highest energy 
non--thermal particles; for those with energy $\approx E_u$, it can be 
shown that the ratio of synchrotron deceleration time to shell crossing
time is \cite{vietri}:
\begin{equation}
\frac{t_{sy}}{t_{cr}} = 90.0 n_1^{4/3} E_{51}^{1/3} \xi^{-3/2}\;,
\end{equation}
which depends weakly upon all parameters except the efficiency $\xi$ 
with which equipartition magnetic fields are built up behind the shocks.

The total energy loss through synchrotron emission by non--thermal protons
is obtained by integrating the energy loss rate $\dot{\epsilon}$ over the
particle spectral distribution, with the normalization given above, and
multiplying times the flight time across the shell thickness, which is the
region over which the magnetic field is appreciable. Transforming then to
the Earth frame, the total energy radiated is
\begin{equation}
\label{fluence}
E_{sy} = 3.0\times10^{49}\; erg \; 
\eta^{-1/3} n_1^{-7/6} E_{51}^{5/3} \xi^{3/2}\;.
\end{equation}
For the favoured values $\eta=10^3$ and $E_{51} = 4$, I find $E_{sy} \approx
3.0\times 10^{49} \; erg$. For $p = 2.3$, the limit of the range allowed by 
fitting the UHECRs' spectrum at Earth \cite{wax3}, the above computation yields
$E_{sy} \approx 3\times10^{48} \; erg$. The reduction of the yield in $GeV$ 
photons is due to the fact that steep particle spectra have less energy in 
the high--energy region, where synchrotron losses are stronger. In summary,
I expect a fraction $\approx 0.01$ of the total photon energy release to
end up in the $GeV$ region. 

The spectrum can now be rewritten in terms of the photon
energy as observed at the Earth $E_{ph}$ in units of the cutoff energy,
Eq. \ref{cutoff}, $x \equiv E_{ph}/E_{co}$, as
\begin{equation}
\label{spectrum}
d\!E_{sy}^{(sp)} = \frac{E_{sy}}{s} x^{-s} \; d\!x
\end{equation}
with $E_{sy}$ given above. Since detectors in this energy range are
obviously photon counting devices, it is convenient to give also the
particle spectrum:
\begin{equation}
\label{particlespectrum}
d\!N_{sy}^{(sp)} = \frac{E_{sy}}{sE_{co}} x^{-(s+1)} \; d\!x \approx
\frac{1.0\times10^{50}}{s} x^{-(s+1)} \; d\!x \;.
\end{equation}

The last loose end left is to check whether the optical depth for pair creation
against the photons emitted by the electrons exceeds unity: in this case, the
$GeV$--range photons would be degraded in energy, and no flux near the photon
cutoff energy might reach the Earth. I follow in this ref. \cite{bah}.
The total optical depth is $\tau_{\gamma\gamma} = r_{sh}/l_{\gamma\gamma}$,
where the mean free path for photon--photon pair creation is 
$l_{\gamma\gamma} = \sigma_T U_\gamma\epsilon_{ph}/(4m_e c^2)^2$. Here
$\sigma_T$ is the Thompson cross section, $U_\gamma = E_{GRB} /(4\pi r_d^3)$
is the photon energy density in the shell frame, and $\epsilon_{ph}$ the test
photon energy also in the shell frame. The above formula approximates the 
photon/photon pair creation cross--section as a constant, $\approx 
3\sigma_T/16$, neglecting its decrease with increasing energies. It is thus,
strictly speaking, an upper limit to the optical depth, which is adequate
here. It can easily be seen with the values provided above that 
\begin{equation}
\tau_{\gamma\gamma} = 1.2\times 10^{-5} E_{51}^{1/3} \eta^{1/3} 
\frac{\epsilon_{ph}}{m_e c^2} \;.
\end{equation}
For the usual favoured values $E_{51} =4, \eta=10^3$, I find that
$\tau_{\gamma\gamma} = 1$ only for photons which, as seen from Earth,
exceed $3\; TeV$. Thus there will be no energy degradation {\it in situ}
because of photon/photon pair creation.

This contrasts with the opposite result obtained in ref.\cite{bah}, but 
this is because they considered the other scenario for the generation of 
the shock, \ie, where two relativistic shells collide with each other. 
In their scenario the shock occurs at smaller radii, so that 
the photon  energy density is much higher than in the scenario adopted here. 
This is the reason why I concentrated on ISM shock scenario. It should
be stressed that UHECRs are accelerated in both scenarios, the only
difference being that no $GeV$ photons can come out of one of them, and that,
potentially, detection of these high energy photons may distinguish 
between the two GRB scenarios. 

\paragraph*{Observability.}

I now consider the observability of the highest energy photons, those with
energies close to $E_{co}$.
High energy photons produce pairs by collisions with photons of the Infra Red 
or Microwave Background, thus losing energy efficiently. This limits the range 
from which $300\; GeV$ photons can reach us to $D_m \approx 300 \; Mpc$ 
\cite{gzk}. The flatness of the $\log N-\log S$ relation for bursts 
\cite{pendleton}
implies that we are already seeing the edge of the GRBs' distribution; thus
the rate of GRBs deduced from observations, $\approx 30 \; yr^{-1} \; 
Gpc^{-3}$, is not likely to be very incomplete. From this, I deduce a rate 
of GRBs inside $D_m$ of $3\; yr^{-1}$. For these distances, the expected 
fluence from Eq. \ref{fluence} is $3\times10^{-6} \; erg \; cm^{-2}$. 

There are currently no experiments which can detect showers initiated by 
primaries with energies around $E_{co}$ with the required sensitivity, but the 
next generation of high--altitude ($> 4000\;m\; a.s.l.$) detectors currently 
being planned will have relevant detection thresholds of $\approx 10\; GeV$. 
Since I argued that the spectrum is expected to be very flat, most energy, 
though not most counts, will be deposited above this detection threshold, so
that the limiting factor will be the experiments' detection surface.
As an example, ARGO \cite{argo} can detect signals down
to flux levels of $\approx 3\times 10^{-7}\; erg \; cm^{-2}$ for
spectra extending out to $300 \; GeV$, provided the bursts last $1\; s$,
and the spectra are flatter than $s\approx 1.5$. However, with this low 
detection threshold, a burst with fluence $3\times10^{-6}\; erg\; cm^{-2}$ could
be detected over background noise even if it lasted $100\; s$.

\paragraph*{Implications.}

High energy photons will be emitted as long as protons are accelerated to very 
high energies, which only requires the presence of large
magnetic fields. There is no obvious reason why the field should decay on the
timescale of the burst. In fact, first, as seen from Earth, the shell remains 
relativistic for about a month after the burst \cite{vietri2}, which implies
that relativistic electrons will be available in the post--shock region
to generate a non--negligible magnetic field. Second, after the burst, the 
electrons exchange energy with protons through a variety of 
processes. So long as electrons are thusly kept relatively 
hot, their chaotic motion may maintain an appreciable
magnetic field. Since processes coupling electrons to
protons are relatively inefficient in transfering energy, 
some magnetic field may be left for some time after the burst. So
observations of the secondary burst discussed in this paper may reveal 
burst durations appreciably longer than those of their lower energy 
counterparts. 

Another important issue is whether this emission may be masked by
Inverse Compton effects: it was shown \cite{rees94} that photons
with energies as high as $10\; (\eta/100)^6 \; GeV$, very incertain because 
of the steep dependence upon $\eta$, can be produced this way.
In this case, the expected spectrum is simply that of lower energy, seed
phtons, which get a kick to higher energies: typically the photon number
spectrum is $\propto \nu^{-2}$, different from Eq. \ref{spectrum}. 
This furthermore implies that the energy released
per photon energy decade is constant, and thus exceeds the estimate
of Eq. \ref{fluence} by about two orders of magnitude. 
Thus, spectral steepness and fluence allow
an easy discrimination between the two mechanisms.

The above argument finds an interesting application at intermediate energies,
$\approx 1\; GeV$. One may in fact wonder whether the low energy tail of
the emission derived in Eq. \ref{spectrum} has been observed already,
since some bursts do show detectable fluxes at these energies \cite{bursts}.
This can be certainly excluded for {\it GB 930131}, which has comparable
fluences in the low--energy BATSE spectrum ($\approx 1 \; MeV$) and in the
$GeV$ region, with nearly identical spectral slopes \cite{kou}. The situation
is however less clearcut for other bursts, like {\it GB 940217, GB 910503},
which have fluences, at their respective highest energies, much lower 
than those emitted at lower energies, and thus in keeping with the
estimate of Eq. \ref{fluence}. It should also be noticed that all these bursts 
show, at the highest energies, burst duration much longer than at low energies, 
a fact naturally accounted for in this model, as discussed above.

A mechanism for the production of
delayed $GeV/TeV$ photons from GRBs based upon the presence of UHECRs has
been proposed \cite{wax4}, but it differs greatly from the present one: there
UHECRs produce photons through photopion processes off CMB photons in flight
from the site of the burst to our detectors, while in this model the
emission mechanism is synchrotron {\it in situ}.

\paragraph*{Summary.}

Physical conditions at the relativistic shocks which 
give origin to GRBs are surely favorable to acceleration of high
energy particles. If the energy in UHECRs is comparable to that released
in photons (and thus if UHECRs from GRBs account for the whole flux
of UHECRs at Earth), the fluxes are those predicted above. Otherwise,
the observed flux of $GeV$ photons will allow measurement of the
fraction of energy channeled into non--thermal particles. 

I have shown that the synchrotron emission spectrum from
protons extends to $E_{co} \approx 300\; GeV$ (Eq. \ref{cutoff}) and should be 
rather flat (Eq. \ref{spectrum}), that the total energy release is $E_{sy} 
\approx 0.01 E_{GRB}$ (Eq. \ref{fluence}), and that a few such events per year
will surely become observable with the next generation of high altitude,
small size air shower detectors. I have also argued that all $GeV$ emission 
observed to date from GRBs is compatible with this model. 

\paragraph*{Acknowledgements.}

Thanks are due to G.C. Perola for a critical reading of the manuscript.

\end{document}